\magnification=1200

\def\a{\alpha}\def\c{\chi}\def\d{\delta}
\def\f{\phi}
\def
\p{\pi}
\def\y{\eta}

\def\L{\Lambda}

\def\na{\nabla}
\def\id{\equiv}\def\mo{{-1}}

\def\dix{\int d^2x\ \sqrt{-g}\ }\def\ds{ds^2=}

\def\fe{field equations }\def\bh{black hole }

\def\bhs{black holes }

\def\sch{Schwarzschild }

\def\GR{general relativity }

\def\section#1{\bigskip\noindent{\bf#1}\smallskip}

\def\PR#1{Phys.\ Rev.\ {\bf#1}}\def\CQG#1{Class.\ Quantum Grav.\ {\bf#1}}
\def\NP#1{Nucl.\ Phys.\ {\bf#1}}
\def\JMP#1{J.\ Math.\ Phys.\ {\bf#1}}

\def\PRep#1{Phys.\ Rep.\ {\bf#1}}
\def\AoP#1#2{Ann.\ Phys.\ {\bf#1}}
\def\hep#1{{\tt hep-th/#1}}

\def\ref#1{\medskip\everypar={\hangindent 2\parindent}#1}
\def\beginref{\begingroup
\bigskip
\centerline{\bf References}
\nobreak\noindent}
\def\endref{\par\endgroup}

\def\kokh{{1+k\over2+k+h}}\def\uokh{{1\over2+k+h}}
\def\bb{Bekenstein bound }\def\gsl{generalized second law }
\def\td{two-dimensional }\def\ehf{e^{-2h\f}}\def\ef{e^{-2\f}}

{\nopagenumbers
\ \vskip60pt
\centerline{\bf Entropy bounds in two dimensions}
\vskip40pt
\centerline{\bf S. Mignemi\footnote{$^\dagger$}{\rm
e-mail: smignemi@unica.it}}
\vskip20pt
\centerline{Dipartimento di Matematica, Universit\`a di Cagliari}
\centerline{viale Merello 92, 09123 Cagliari, Italy}
\centerline{and}
\centerline{INFN, Sezione di Cagliari}
\vskip80pt
\centerline{\bf ABSTRACT}
\smallskip
{\noindent We discuss entropy bounds for a class of two-dimensional
gravity models. While the Bekenstein bound can be proved to hold in
general
for weakly gravitating matter, the analogous of the holographic bound
is not universal, but depends on the specific model considered.}
\vfil\eject}

It was observed by Bekenstein [1], that the
entropy $S$ contained in a weakly gravitating system must satisfy
the bound $S\le2\p ER$,
where $R$ is the linear size and $E$ the energy of the system.
This bound can be deduced from the \gsl of \bh thermodynamics
by means of a gedanken experiment known as  Geroch process,
and is known to be satisfied by all reasonable weakly gravitating
matter systems. In particular, it is saturated by the \sch
black hole. The validity of the \bb for any $D>2$ was later proved
by Bousso [2].

A different bound, which is believed to hold also for strongly
gravitating systems is the holographic bound, $S\le {A\over4}$,
where $A$ is the area of a surface enclosing the system [3]. This
bound is justified by the observation that $A/4$ is the maximum
entropy that can be contained in a region, before the matter
collapses to a black hole.

In this letter we discuss the possibility of extending these
bounds to $D=2$. This is motivated by the need of exploring the
universality of the previous bounds on more general gravitational
theories than \GR. Two-dimensional gravity offers a possibility for
this, since it gives rise to several different models with varied
physical properties.
Moreover, in 2D the holographic bound cannot be defined in terms of
area, since in that case the horizon of a \bh reduces to a point.
Also, in ref. [4], it was observed that, in contrast with higher
dimensions, two-dimensional \bhs in general do not satisfy the \bb.
This fact was explained making recourse to the invariance of
some two-dimensional models under dilatations.
We shall prove that in any case, if one restricts to weakly gravitating
matter, the \bb holds and is a universal property
independent of the specific model.
On the contrary, we show that a bound of holographic type does
not hold in general, but is necessarily model dependent.

\bigbreak
It is well known that gravity cannot be introduced in a unique
way in two dimensions. This is due to the fact that the
Einstein action is trivial in 2D, and therefore one is forced
to introduce a scalar field $\y$ (dilaton) in order to write an action
principle for \td gravity. This field admits an arbitrary
potential, which gives rise to an infinity of different models [5].

In the following we shall consider for definiteness
a special class of these models, with action
$$I=\dix\ef[R-4k(\na\f)^2+\L\ehf],\eqno(1)$$
where $k$ and $h$ are two arbitrary parameters. In order to
have well defined \bh solutions, we require that
$k>-1$, $h>-(k+2)$.
The action (1) is rather general and includes as special cases many
of the best known models. For $h=0$ it reduces to the models
studied in [6], while for $k=0$ gives rise to the models of [7].
The case $h=0$, $k=0$ is the JT model [8].

The \fe derived from (1) admit the solution
$$\ds-V(x)\,dt^2+V^\mo(x)\,dx^2,\qquad\qquad \ef=e^{-2\f_0}\,
|x|^{1\over1+k},\eqno(2)$$
where
$$V(x)={(1+k)^2\over2+k+h}\left(\L\,|x|^{2+2k+h\over1+k}-m\,|x|^{k\over1+k}
\right),\eqno(3)$$
and $m$ is a parameter.
For positive $\L$ and $m$, these solutions describe \bhs with a horizon at
$$x_0=\left(m\over\L\right)^\kokh.$$
The constant $e^{-2\f_0}$ is arbitrary and its inverse corresponds to
the \td Newton constant. We shall set it to 1 without loss of
generality. It is evident that varying the parameters $h$ and $k$
one can obtain a large variety of models with rather different physical
properties.

The mass $M$, temperature $T$ and entropy $S$ of the \bhs (2)-(3)
can be defined as in [9], and are are given by
$$M=\kokh\, {m\over2},\qquad T={1+k\over4\p}\left(\L\,m^{1+k+h}\right)^\uokh,
\qquad S=2\p\left(m\over\L\right)^\uokh.\eqno(4)$$

If one identifies the size of the \bh $R$ with $x_0$ and its
energy $E$ with $M$, it results that
$$2\p ER\propto m^{3+2k+h\over2+k+h}\propto S^{3+2k+h},$$
and clearly one can always violate the Bekenstein bound.

This can be seen as a consequence of the unusual thermodynamical
properties of 2D black holes. In fact, splitting a \td\bh into
\bhs of smaller mass can be entropically favoured [10,7].
This is evident from the expression (4) for the entropy, if $k+h>-1$.

For our considerations, it is also instructive to  write the entropy
as a function of the \bh length, as
$$S=2\p x_0^{1\over 1+k}.\eqno(5)$$
If $k>0$, the entropy of $N$ black holes contained in a given
portion of space of size $R$ is greater than that of a single
black hole of the same size by a factor $N^{k\over1+k}$.
Hence, by increasing the number $N$ of \bhs one can apparently store an
arbitrary amount of entropy in a given region.
This behaviour may be seen as a consequence of the repulsive nature of the
potential (3) at short distances for positive $k$. It must
be noted however that in this case the system of $N$ \bhs is unstable, since the
\bhs will tend to move away from each other. As we shall show, in
order to obtain the correct entropy bounds, one must therefore take
into account the interaction between the $N$ black holes.

Although in 2D the Bekenstein bound is not valid for a collection of
\bhs, it still holds for weakly gravitating matter.
This can be shown by slightly modifying the proof based on
the Geroch process given in ref.\ [2], in such a way to avoid any
reference to the \bh area. The proof relies on classical \GR
and the \gsl and neglects quantum effect [11].

Consider a weakly gravitating system of total energy $E$ and
linear size $R$. Move the system towards a \bh of radius much
greater than $R$, described by a metric in the \sch gauge (2),
with arbitrary $V(x)$ having a simple zero at $x=x_0$.
Lower slowly the system until it is just outside the horizon and
finally drop it in.

The mass added to the \bh is given by the energy $E$ of the
system, redshifted according to the position of the center of mass
at the drop-off point, i.e. at a distance $R$ from the horizon.
Near the horizon the metric can be approximated as
$$V(x)\approx{dV\over dx}\Big|_{x=x_0}(x-x_0)=4\p T(x-x_0),$$
where we have used the standard definition of temperature $T={1\over4\p}
{dV\over dx}\big|_{x=x_0}$.
Denoting $y=x-x_0$, the proper distance $l$  from the horizon is
given by
$$l=\int{dy\over\sqrt{4\p Ty}}=2\sqrt{y\over4\p T}.$$
It follows that the redshift factor is
$$\c(l)\id\sqrt{V(l)}\approx2\p Tl.$$
The mass added to the \bh is therefore
$$\d M=E\c(l)\Big|_R=2\p TR,$$
and the black hole entropy increases by
$$\d S={dS\over dM}\d M= {\d M\over T}= 2\p ER,$$
where we have used the thermodynamical relation $T^\mo=dS/dM$.
By the \gsl this increase must at least compensate for the lost
matter entropy, and hence
$$S_{matt}\le2\p ER.$$

This proof is valid in any dimensions $D\ge2$ and is independent
of the specific action or solution, being based uniquely on the
definition of a horizon and of the thermodynamical quantities and
the generalized second law.

This is an indication of the universality of the Bekenstein bound
and may be a sign of its independence from other kind
of entropy bounds such that the holographic bound, which in 2D cannot
be formulated straightforwardly and is model-dependent.

Indeed, let us now try to define a \td entropy bound which is
valid also for strongly gravitating systems, as black holes.
Of course, a holographic bound cannot be defined as in higher
dimensions, due to the lack of an area in 2D.
However, in analogy to higher dimensions, one can attempt to define
a bound as the minimum entropy necessary for the formation of
a black hole in a region of size $R$.
In fact, if the entropy of the matter contained in a given region
exceeded that of a \bh of the same size, adding matter till the
formation of a black hole would violate the \gsl.

From (5), one gets therefore, for a single black hole,
$$S_{matt}\le2\p R^{1\over1+k}.$$

One may object that, as discussed above, for positive $k$ it is
possible to store $N$ \bhs in the same amount of space, which
would therefore contain a larger amount of entropy.
However, these \bhs repel each other, and such configuration cannot
be stable. Neglecting the backreaction, we may assume that a stable
configuration can be approximated
when the distance between the \bhs minimizes the potential $V(x)$.
For a \bh of mass $m/N$, $V$ has a minimum at
$$x_m=\left({k\over2+2k+h}\ {m\over N\L}\right)^\kokh.$$
Hence, a system of $N$ \bhs of mass $m/N$ in equilibrium will have
size $R\approx Nx_m$ and entropy
$$S\approx2\p\left(1+{2+k+h\over k}\right)^\uokh R^{1\over1+k}.$$
It follows that also in this case one can introduce a bound
$S_{matt}<2\p\a R^{1\over1+k}$, for a given constant $\a$.
\smallskip
The results of this letter seem to support the view that while the
\bb is inherent to the definition of \bh thermodynamics in any
metric theory of gravity, the existence of a holographic bound
depends also on the dynamics of the specific model of gravity.
This is not in contrast with the holographic principle [3], since
it is reasonable to assume that the boundary dynamics depends on
the theory that governs the bulk.
\bigskip

\beginref
\ref [1] J. Bekenstein, \PR{D23}, 287 (1981).
\ref [2] R. Bousso, JHEP\ {\bf 0104}, 035 (2001).
\ref [3] G. 't Hooft, \hep{9310026}; L. Susskind, \JMP{36}, 6377
(1995).
\ref [4] M. Cadoni, P. Carta and S. Mignemi, \NP{B632}, 383 (2002).
\ref [5] For a recent review, see D. Grumiller, W. Kummer and
D.V. Vassilevich, \PRep{369}, 327 (2002).
\ref [6] J.P.L. Lemos and P.M. S\'a, \PR{D49}, 2997 (1994);
M. Cadoni and S. Mignemi, \NP{B427}, 669 (1994).
\ref [7] S. Mignemi,  \PR{D50}, 4733 (1994); \AoP{245}, 23 (1996).
\ref [8] C. Teitelboim, in {\sl Quantum Theory of gravity}, S.M. Christensen,
ed.\ (Adam Hilger, Bristol, 1984); R. Jackiw, {\sl ibidem}.
\ref [9] V.P. Frolov, \PR{D46}, 5383 (1992).
\ref [10] R.B. Mann and T.G. Steele, \CQG{9}, 475 (1992);
P.T. Landsberg and R.B. Mann, \CQG{10}, 2373 (1993).
\ref [11] W.G. Unruh and R.M. Wald, \PR{D25}, 942 (1982).

\endref

 \end